\documentstyle[11pt]{article}
\def\sub#1{_{_{#1}}}
\begin{document}
\baselineskip=14pt
\title{\bf Gravitational Screening}
\author{E. A. Spiegel \\
{\it Department of Astronomy} \\
{\it Columbia University} \\
{\it New York, NY 10027 USA}\\ }
\date{October 2, 1997}
\maketitle
\begin{abstract}
\baselineskip=12pt
\noindent
Calculations of the stopping power of a medium lead to divergent
integrals in gravitational theory as they do in the analogous
electromagnetic problem.  In the em case, one 
introduces (normally in an arbitrary manner)
the Debye length as a cutoff at large distances to remove
the divergence.  We show here that in the Newtonian gravitational
analogue the distant cutoff problem is solved by including the
self-gravity of the medium.  Then the Jeans length appears naturally as a
cutoff.  Despite the different sign of the coupling constant from the
case of the electric plasma, the same sort of theory works in both
cases and removes the need of introducing {\it ad hoc} cutoffs from both
of them.
\end{abstract}
\baselineskip=12pt
\section{Gravitational Stopping Power} 
In calculating the total Coulomb scattering cross section, one is
normally faced at the end with an integral over all scattering angles.
This integral diverges and some fix is needed.  In plasma physics one
usually says that the smallest scattering occurs at the largest impact
parameter and that particles at sufficiently great distances do not feel
the scatterer.  The reason given is that the intervening medium
effectively screens the electric potential for any object passing at a
distance greater than the Debye length.  Another example of this problem
is seen in the calculation of the drag of a medium on a charged object
passing through it.  Again the Debye-H\"uckel \cite{Debye} theory is
called to the rescue.

In this discussion I am concerned with the second example but mainly
in the gravitational setting.  The problem is how to deal with
the divergence of the integral for the gravitational drag on a massive
object passing through a uniform medium.  A cutoff is needed but the
question is whether we can simply use the seemingly natural Jeans
length as a cutoff in the purely gravitational case.  As we shall see,
if one takes into account explicitly the self-gravity of the ambient
medium, there is no need to invoke an {\it ad hoc} assumption about
cutoffs.  The problem solves itself, and the same is true for a
charged projectile. There does remain the need for a short range
cutoff, but this has to do with processes that are not closely
connected with the present considerations.

The discussion here will be couched in purely Newtonian terms but  I
feel that this needs no apology.  After all, this volume has been
assembled to honor Engelbert Schucking, one of the great heroes of the
neoNewtonian revolution in cosmology that began in the mid-thirties.
What better precedent could one have?  In fact, Engelbert is
indirectly responsible for the work reported here.

After I had attended the first Texas Conference in Relativistic
Astrophysics that Engelbert and his fellow Texans organized in 1964, I
returned with ideas of how to speed up the gravitational collapse of the
core of a galaxy (a problem which I had been discussing previously with
M.A. Ruderman).  I wrote out my notions and sent them to D.E. Osterbrock
who, in a pleasant way, raised an objection that I never overcame.
However, the following calculations arose from my attempts to respond to
him.  My understanding of what it all meant was advanced by comments of
C.W.  Misner.  So with thanks to Engelbert whose actions got me onto this
path of agreeable interactions, both personal and gravitational, let me
next state the problem that arises when the self-gravity of the ambient
medium is not included.
\section{A Drag Crisis} 
When an object of mass $m$ moves with velocity ${\bf U}$ through an
infinite, uniform medium of density $\rho_0$,  the fluid, initially at
rest, begins to move with velocity ${\bf u}$ and its density is modified
to $\rho$.  The fluid dynamics is then described by these equations: $$
\partial_t (\rho {\bf u}) + \nabla\cdot(\rho {\bf u} {\bf u})
= - \nabla p - \rho \nabla V \eqno(2.1) $$  $$
\partial_t \rho + \nabla\cdot(\rho {\bf u}) = 0 \eqno(2.2) $$
where $V$ is the potential of the gravitational field.  The field is,
in this section {\it only}, assumed to come from the passing
object alone, so that  $$
\Delta V = 4\pi G m \delta({\bf x} - {\bf U}t) \ , \eqno(2.3) $$
where $\Delta$ is the Laplacian and $\delta$ is the Dirac function.
We assume that the drag on the object is so weak that we may
treat ${\bf U}$ as constant.

Let us write $\rho = \rho_0(1+\psi)$ and assume that $|\psi|<<1$
everywhere with similar assumptions about the fluid velocity and
pressure.  Then we obtain linear equations for the perturbation
caused by the intruding object.  Though there are also some
results on the nonlinear problem, I will restrict myself to the
linear case here since this illustrates nicely the point I want
to raise.

With the assumption that $p=p(\rho)$ we can (in the linear case)
write the pressure perturbation as $a^2 \rho_0 \psi$, where $a$
is the adiabatic speed of sound, given by
$a^2 = [\partial p/\partial \rho]_0$.  It is then a straightforward
matter to linearize the equations and boil them down to an equation
for $\psi$.  The linearized forms of (2.1) and (2.2) are $$
\partial_t {\bf u} = a^2 \nabla \psi + \nabla V \eqno(2.4) $$
$$\partial_t \psi =  - \nabla\cdot {\bf u}. \eqno(2.5) $$
We are considering the potential produced by the incoming object to
be small also, so that the full term $\nabla V$ appears in (2.4)
With the help of (2.3), we may condense this to an inhomogeneous
wave equation for $\psi \,$: $$
\partial_t^2 \psi - a^2 \Delta \psi = 4\pi G m \delta({\bf x} - {\bf
U}t) \ . \eqno(2.6) $$

The flow induced by the object is called an accretion flow though,
to study the actual accretion, one might also include
a sink term on the right of (2.2).  What I am interested in here
is the so-called stopping power of the medium.  In the next section
we shall calculate this by evaluating the gravitational force of the
disturbed medium on the object.  This can be done as described
by Landau and Lifshitz \cite{LL} for the standard Cherenkov problem.
I postpone this to the next section since I want to explain here why
something extra is needed.  That something is the self-force of the
medium, which I am leaving out in this section.  The problem without
self-interaction has a significant literature.

The solution to the drag problem in the gravitational case (with magnetic
field included) was first published by Dokuchaev \cite{Dok}, as I learned
from G. Golytsin.  F.D. Kahn (private communication) and K.H.  Prendergast
(private communication) also were early among those who thought about this
calculation.  A very simple way of solving (2.6) is to be found in the
book of Ward \cite{Ward} as S.  Childress has told me.  There are also
papers dealing with the wakes of plasma probes \cite{Lam}, charged
satellites \cite{Kraus}, and massive objects in galactic disks
\cite{Jul} \cite{Sim}.  And of course, behind and beyond all this is the
heritage of the nuclear and atomic literature on stopping power with
contributions from Bethe, Bohr, Fermi and other Olympians (see
\cite{Ichi}).

The line along which the object moves is called the accretion axis
\cite{Lytt}.  In the reference frame of the object, we may find
solutions which are steady and symmetric around the accretion axis
\cite{Spieg} by the methods of linear acoustic theory \cite{Ward}. To
express these, we let $U = |{\bf U}|$ and introduce the Mach number, $M =
U/a$, and the accretion radius, $$
R\sub A = {2Gm\over a^2} \ . \eqno(2.7)$$
We also let $\theta$ be the angle measured from the downstream accretion
radius and $r$ be the distance from the object, both in the object frame.
Then, for subsonic motion ($M < 1$), we obtain the solution $$
\psi = {R\sub A M^2 \over r \sqrt{(1 - M^2 \sin^2 \theta)}} \ .
\eqno(2.8) $$
For supersonic motion ($M > 1$) the disturbance is confined to the region
inside the downstream Mach cone; that is, $\psi = 0$ for $\theta > \arcsin
M^{-1}$.  Inside the downstream Mach cone, the solution is again (2.8).
These solutions may be verified by direct substitution into (2.6).

Apart from remarking that the shock wave at $\theta =
\arcsin M^{-1}$ is singular on account of the linearization and
the point nature of the interloper, I forgo discussion of this solution
(but see \cite{Spieg} for some illustrations).  What is of concern here
are two conclusions about the drag force that may be drawn from the
formulation of the problem as stated.  This force is, on grounds of
symmetry, along the accretion axis, and is written in the form
${\bf F} = - F\sub A {\bf U}/U$ where, on dimensional grounds, the
accretion drag is $$
F\sub A  = \pi R^2\sub A \rho_0 U^2 C\sub A \ . \eqno(2.9) $$
Here, we have introduced the accretion drag coefficient $C\sub A$ that
contains the crux of the problem.

The first conclusion is that the drag in the subsonic case, $M<1$, is
zero.  This is a familiar result in terrestrial subsonic acoustics
since, there too, subsonic flows have fore-aft symmetry.  The second
conclusion is that the drag formula makes no sense in the supersonic case
unless we introduce some additional physics.  To these rather bald
statements, let me add some heuristic justification for those who may
not wish to await the more extensive calculations of the next section.
Those who do not care much for heuristic arguments might skip to the
next section at once.

As in the Coulombic case, one may make the sudden approximation
to calculate the stopping power \cite{Rud}.  A fluid particle going by
in the object's rest frame is given a transverse kick and starts moving
with velocity $v_\bot$ toward the accretion axis.  This velocity is
really a transverse momentum per unit mass and we may therefore estimate
it as the force (per unit mass) on the fluid element multiplied by the
time it acts.  That force per unit mass is $Gm/b^2$, where $b$ is the
impact parameter.  The time of action is approximately $2b/U$.  The
product is then $$
v_\bot = {2Gm\over bU} \ .  \eqno(2.10)  $$

Energy is delivered to the medium at the rate ${\bf F\cdot U}$.  To
estimate this, we note that the kinetic energy of the medium is created
at the rate ${1\over 2} \rho_0 v_\bot^2 U$ per unit time per unit of area
normal to the direction of symmetry axis. If we integrate over that area,
we obtain the rate of energy delivered to the medium (or lost by the
object).  On dividing that by $U$, we find that the force has magnitude $$
F\sub A = 2 \pi \int_{b_{min}}^{b_{max}} {1\over 2}\rho_0 v_\bot^2 b\,  db
\ ,
\eqno(2.11) $$
where the integration has been limited between a minimum and a maximum
impact parameter.  If we put in the derived estimate for $v_\bot$, and
compare the result with (2.9), we find that the accretion drag coefficient
of (2.9) is $$
C\sub A = \log{b_{max}\over b_{min}}\ . \eqno(2.12) $$

The cutoff impact parameters have here been put in by hand so as to
bring out the nature of the problem.  To get a finite
answer we need to see where the present formulation breaks down and
either cut off there or introduce some additional physics.  For the inner
cutoff, one possibility is that we may acknowledge that the object
actually has a finite radius, $R$. Then, as Eddington showed
\cite{Edd1}, a fluid element coming in with impact parameter $b<b_0$,
where $$ b_0 = \sqrt{R(R\sub A+R)} \ , \eqno(2.13) $$
will hit the object.  Assuming that the collision is sticky, we can
make $b_0$ the inner cutoff.  Of course, we should add a correction
to the drag to account for the particles that hit the object, but
I will not do that here.  Our concern is really with the outer
cutoff.

If the medium is expanding, that will produce a cutoff at a distance
where the Hubble velocity, $H_0r$, is comparable to $v_\bot$.  With the
expression just derived, $v_\bot = 2Gm/(Ub)$, this leads us a value $$
b_{max}^2 = {R\sub A U\over H_0} = \sqrt{R\sub A R\sub U} \ , \eqno(2.14)
$$
where $R\sub U=U/H_0$ is the distance at which expansion speed is the same
as that of the object with respect to the medium.

However, the real issue here is not so much the value of the drag but
its interest in bringing out the nature of the gravitational behavior
of a fluid.  In that case we want to know how faithful the analogy to the
electromagnetic case is.  What we have seen here is that a cutoff is
needed.  The question is whether the Jeans length $\pi a/\sqrt{4\pi G
\rho_0}$ may be used for this cutoff if extraneous phenomena do not
intervene first.  A positive answer is proposed in the next section.
\section{Saved by Self-Gravity} 
In this section I sketch the calculation of the drag with
the inclusion of the self-gravity of the medium.  Apart from that
addition, the approach is along conventional lines such as that
of Landau and Lifshitz, \cite{LL}.  The effect of self-gravity
is to invalidate both of the conclusions about the drag mentioned in the
previous section.  When I first told these things to knowledgeable people
I found that some of them did not want accept one or the other of these
results.  Martin Lampe, who was sympathetic to my problem,
independently repeated the calculations and verified the revised
conclusions in what I regard as an act of remarkable kindness, even for
someone who might be fond of contour integrations.

To include the effect of the self-gravity of the medium, we add the
term $4\pi G \rho$ to the right hand side of equation (2.4).  This however
introduces a problem: there is no longer a static homogeneous
solution to the equations, even in the absence of the intruding object.
This difficulty is one of long standing and there have been two general
ways of coping with it, if we leave aside the Jeans \cite{Jeans} approach
of pretending it isn't there.  One thing to do is to give up the
simplicity of the homogeneous background and look for inhomogeneous
equilibria.  In many such structures, the gravitational force balances the
pressure gradient, and so their sizes are on the order of the Jeans
length, which in this way can also get into the act.   This is a less
interesting direction for the present discussion than the second one,
which is to modify the theory to allow the existence of a static,
homogenous and infinite medium.  Let us take the latter route.

In the analogous case of plasma dynamics, the physics itself provides the
solution.  There, in first approximation, $\rho$ is the density of the
electron gas and it is proportional to the charge density.  If the medium
is to be electrically neutral, one must include the positive charge
density of the ions on the right hand side of the Poisson equation.  In
the simplest models, this ion charge density is a constant but, in more
sophisticated versions, it has a dynamics that is excited through
coupling to the electrons through their joint potential. One then has a
two-fluid problem.

In the gravitational case, we may similarly modify the Poisson equation,
(2.4), to $$
\Delta V = 4\pi G m \delta({\bf x} - {\bf U}t) + 4\pi G (\rho -
\rho_{\Lambda}) \ . \eqno(3.1) $$
where $\rho_\Lambda$ is a (constant) density of negative gravitational
mass.  This is the approach I will follow here but, in passing, I feel it
is worth asking whether one should go all the way in the analogy to the
plasma dynamics and couple in some dynamical model for this dark
antigravitational fluid to see where that leads.

As is quite clear, we can absorb the $\rho_\Lambda$ term into
the potential with the replacement $V\rightarrow V + (4\pi/6) G
\rho_\Lambda r^2$.  Then, the $\rho_\Lambda$ disappears from the
Poisson equation and reappears explicitly on the right hand side
of the equation of motion, (2.1), in the form ${1\over 3} \Lambda
\rho {\bf r}$, where $\Lambda = 4\pi G \rho_\Lambda$.  This is the
Newtonian analogue of Einstein's cosmological term.  According
to Bondi \cite{Bondi}, the cosmological term was introduced in the
Newtonian setting by Seeliger in 1895.  Engelbert tells me that Seeliger
in fact put in a shielding effect, not a repulsive force, and that
Einstein believed that he too was doing such a thing in the relativistic
case.  However, Engelbert remarks, it was Eddington \cite{Edd2}
who first noted that the cosmological term represented a repulsive force.
I believe that its use is far preferable to Jeans' disregard of the
problem of finding a static homogenous solution, especially as (3.1) is so
suggestive.

Now we may proceed as before with the linear theory.  We again let
$\rho = \rho_0(1+\psi)$ where, this time, $\rho_0=\rho_\Lambda$.
Thus the introduction of the cosmological density changes nothing that
matters here but it does permit us to expand about a uniform state
in a better frame of mind than Jeans must have been in when he studied
gravitational instability.  On assuming that $|\psi|<<1$ as before,
we obtain an inhomogeneous Klein-Gordon equation (2.6), namely,
$$
\partial_t^2 \psi - a^2 \Delta \psi - a^2 k_J^2 \psi = 4\pi G m
\delta({\bf x} - {\bf U}t) \ , \eqno(3.2) $$
where $k_J^2=4\pi G \rho_0/a^2$ and where $\psi$ is real.

The solution to (3.2) is a linear superposition of the forced, or
driven solution, and of the solution of the homogeneous equation
consisting of the free waves.  The latter, which is needed to solve
the general initial value problem, contains the seeds of gravitational
instability.  In this problem we may expect  this instability to
produce gravitationally condensing wakes that resemble what is
seen in a cloud chamber in the wake of a charged object.  Such a process
was invoked in the steady state theory to produce new galaxies.  I have
not seen any nonlinear treatment of this phenomenon and there is not much
point in pursuing that line without one.  I will therefore adopt initial
conditions in which the free modes are not excited.

To find the forced solution of this problem, we may follow Landau,
Lifshitz and many others into Fourier space, setting $$
\psi({\bf x},t) = \int d^3{\bf k} \, e^{i {\bf k \cdot x}}  \hat \psi
({\bf k},t) \ . \eqno(3.3) $$
Then we take the Fourier transform of (3.3), use $$
\delta({\bf x}-{\bf U}t) = {1\over (2\pi)^3} \int d^3{\bf k}\;
e^{i{\bf k}\cdot ({\bf x}-{\bf U}t)} \ \ ,  \eqno(3.4)$$
and obtain
$$
\partial_t^2 \hat \psi + a^2({\bf k}^2 - k_J^2) \hat \psi
= {G m \over 2 \pi^2}\; e^{-i {\bf k\cdot U}t}\ \ .  \eqno(3.5) $$

At this point, it is possible to anticipate that when we do the
Fourier inversions we shall have to pay attention to poles in
$\hat \psi$.  When these are on the real axis, we shall
have to decide which way to go around them.  To make this choice
unambiguously, we replace $m$ by $m\exp (\epsilon t)$ where $\epsilon$
is an arbitrarily small, positive quantity.  With the introduction
of this standard device, we see that back at $t = - \infty$ there
was no intruding object so that the disturbance is turned on very
gently without making waves.

With $\epsilon=0^+$, we introduce
$$ \omega = {\bf k\cdot U} + i \epsilon. \eqno(3.6) $$
and rewrite (3.5) as
$$
\partial_t^2 \hat \psi + a^2({\bf k}^2 - k_J^2) \hat \psi
= {G m \over 2 \pi^2} e^{-i \omega t}.  \eqno(3.7) $$
The slow turn on lets us concentrate on the forced solution to
(3.6), which is $$
\hat \psi ({\bf k}, t) = {Gm\over 2 \pi^2}\, {e^{-i\omega t} \over
a^2({\bf k}^2 - k_J^2) - \omega^2} \ . \eqno(3.8) $$
For $k_J = 0$, Fourier inversion of this result gives (2.8).

The drag on the object, ${\bf F}$, is the gravitational force
of the medium on the object:
$${\bf F}=-F\sub A{\bf U}/U \ = - m \nabla V_d \; , \eqno(3.9) $$
where $V_d$ is the gravitational potential of the disturbance in the
medium.  The third member of (3.9) is to be evaluated {\it at the
object}.

We obtain $V_d$ from the Poisson equation,
$$
\Delta V_d = 4\pi G \rho_0 \psi\ \ , \eqno(3.10) $$
which we solve by taking its Fourier transform:  $$
\hat V_d = - {4\pi G \rho_0\over {\bf k}^2 } \hat \psi \ . \eqno(3.11) $$
Then, we make a Fourier inversion of (3.10) and take its gradient,
bringing an $i{\bf k}$ down into the integrand.  With the $z$-axis chosen
as the accretion axis, we must evaluate the resulting expression on
the object, which is at $x = 0, y = 0, z = Ut$.  Then, on multiplying
by $m$ and introducing (3.8), we obtain for the accretion
drag force,  $$
{\bf F} = {2i G^2 m^2 \rho_0\over \pi} \int {{\bf k}
e^{\epsilon t}\over {\bf k}^2 [a^2({\bf k}^2 - k_J^2) - \omega^2]}\;
d^3{\bf k}\ .
\eqno(3.12) $$

Multiplication of (3.12) scalarly by ${\bf U}$ and comparison with
(2.9) leads is to the following expression for the accretion drag
for $\epsilon$ going to zero and $t$ finite: $$
F\sub A = {iR\sub A^2\; \rho_0\; U^4 \over 2\pi}\int
{k_z d^3{\bf k}\over {\bf k}^2[a^2({\bf k}^2 - k_J^2) - (k_zU)^2 -
2i\epsilon U k_z]} \ , \eqno(3.13)  $$
where $R\sub A$ is defined in (2.7) and we recall that $M=U/a$.

Let ${\bf q}$ be the component of ${\bf k}$ normal to the accretion
axis so that ${\bf k}=({\bf q},k_z)$.  Then this last result can be
rewritten as $$
F\sub A = iR_A^2\; \rho_0\; M^2 U^2
\int_0^\infty q \; d q \; {\cal I}(q^2) \ ,
\eqno(3.14) $$
where $$
{\cal I}(q^2) = \int_{-\infty}^{+\infty} {k_z\; dk_z\over
(k_z^2 + q^2)[(1-M^2)k_z^2 + q^2 - k_J^2 - 2i\epsilon Mk_z/a]} \ .
\eqno(3.15) $$

When $(q^2-k_J^2)/(1-M^2)>0$, the integrand in (3.15) has four poles
on the imaginary $k_z$-axis and we can safely set $\epsilon=0$.  The
integrand is then seen to be odd and nonsingular and we find ${\cal I} =
0$ in that case.  Therefore, for $M<1$, we have that ${\cal I}=0$ for
$q^2<k_J^2$ while for $M>1$, we find ${\cal I}=0$ for $q^2>k_J^2$.

In the case with the other sign, $(q^2-k_J^2)/(1-M^2)>0$, there are two
poles on the imaginary axis and two poles close to the real axis,
displaced from it by an amount of order $\epsilon$ and in a direction
determined by the sign of $1-M^2$.  We close the contour in the half-plane
away from these latter poles and , for $(q^2-k_j^2)/(1-M^2)<0$, use the
method of residues to find
$$ {\cal I} = {\pi i \over (k_J^2 - q^2 M^2)} {\rm sgn}(1-M^2)
\ .
\eqno(3.16) $$

For the subsonic case, we see that the integral over $q$ has a
nonzero integrand only for $0<q<k_J$ and that the drag is $$
F\sub A = - \pi R\sub A^2 \; \rho_0\ U^2 \; \log{1 \over \sqrt{1-M^2}}\  ,
\ \ \ \ \ \ \ M < 1 \ ,
\eqno(3.17) $$
where, by our sign convention, a negative $F\sub A$ means acceleration.

In the supersonic case, the integral over $q$ is now from $k_J$ to
$\infty$.  The limit of integration at infinity corresponds to small
scales and we shall assume a cutoff at a large wavenumber $q_0$, say.  We
could, for example take $q_0=1/b_0$ (see equation (2.13)) or use $1/R$; I
shall not try to deal with this issue here.  The key point is that there
is again no need for a cutoff at small wavenumbers.  The drag is in this
case is given by
$$
F\sub A = \pi R\sub A^2\; \rho_0 U^2 \;
\log{\sqrt{(Mq_0/k\sub J)^2 - 1\over
M^2 - 1}} \ , \ \ \ \ \ \ \ M < 1 \ ,
\eqno(3.18) $$
where the specification of the large $q_0$ will depend on the problem.

The way in which the introduction of a cutoff is averted by the inclusion
of self-gravity is different in the subsonic and supersonic cases but,
in each case, the usual problem is removed by the presence of the
$k^2\sub A$ term.  There may still be a worry in the divergence of the
drag at Mach one, but this cannot be helped by a cutoff. This
problem may connected with linearization.  Thus, though there is much
to be resolved in this subject, I would claim that the artificial
cutting off the integrals is not needed.
\section{The Message is: The Medium} 
The notion of shielding in an electric plasma is so familiar that one
takes for granted a cutoff at the Debye distance.  However, the
introduction of a cutoff in a gravitational problem may produce
raised eyebrows.  In retrospect, I find the ready acceptance of the {\it
ad hoc} cutoff in the former case as questionable as it would be in the
latter.  Once we have introduced the background cosmological density, the
two situations are {\it almost} identical.  In fact, the only role of the
background density is to allow us to study a homogeneous medium.  Why then
should the electric case be more exempt from careful scrutiny than the
gravitational case?  I believe that, in both cases, although the
introduction of the natural looking cutoffs does seem to gibe with
physical intuition, the argument that often is offered for it is not
complete.

In the example studied here, there is a steady solution in the reference
frame of the intruding object (albeit an unstable one in the gravitational
case).  If we ask what force may be contributed by a Fourier component
of the density perturbation, we may expect that, over a distance of many
wavelengths, the positive and negative density contributions cancel.  On
the other hand, at any distance, there are wavelenths to match and to
exceed that distance.  These contributions are not self-cancelling, but
they may perhaps be weak.  It appears though that they are strong enough
to integrate up to a logarithm and this is why the cutoffs are sometimes
introduced.  In fact, what is suggested by the calculations just outlined
is that that argument is not enough.  It appears that the reason that the
long wavelengths are excluded is that they are evanescent and cannot reach
out over great distances.  Without that factor, the drag does indeed
diverge.  This is not the whole story as we see from the trouble at
Mach one, but I suggest that in introducing cutoffs we need to invoke
some such mechanism.

The same calculations as those described in the previous section can be
done for the one-component electric plasma.  If you calculate the
Coulombic drag, as has been done here for the gravitational case, and
include the self-interaction of the plasma, you find a finite answer with
no ad hockery.  There are some slight differences in the outcome --- in
particular the drag is zero in the subsonic case for a charged object ---
but the removal of the singularity takes place nicely.  It is somewhat
pleasing then to see the gravitational theory offer something to the
study of plasmas since one gets the impression that the flow has been in
the other direction in recent years \cite{Frid}, at least in the
nonrelativistic case.  As to the treatment of such issues in relativity,
we may hope that this will all have been worked out nicely for Engelbert's
120th birthday.

{}

\end{document}